\documentclass[11pt,a4paper]{article}

\usepackage[utf8]{inputenc}
\usepackage[T1]{fontenc}
\usepackage[english]{babel}
\usepackage{lmodern}
\usepackage[margin=1in]{geometry}
\usepackage{microtype}
\usepackage{xcolor}
\usepackage{graphicx}
\usepackage{booktabs}
\usepackage{amsmath,amssymb,amsfonts,mathtools}
\usepackage{braket}
\usepackage{enumitem}
\usepackage{tcolorbox}
\usepackage{titlesec}
\usepackage[colorlinks=true,citecolor=kipublue,linkcolor=kipublue,urlcolor=kipublue]{hyperref}

\definecolor{kipublue}{HTML}{1B3A6B}
\definecolor{kipuaccent}{HTML}{0E8FB3}
\definecolor{kipulight}{HTML}{F2F6FA}

\tcbuselibrary{skins,breakable}

\newtcolorbox{takeaway}{
  enhanced,
  colback=kipulight,
  colframe=kipuaccent,
  boxrule=0.6pt,
  arc=2pt,
  left=10pt,right=10pt,top=8pt,bottom=8pt,
  title={\textbf{Key takeaway}},
  fonttitle=\bfseries\color{white},
  colbacktitle=kipuaccent,
  breakable,
}

\titleformat{\section}{\Large\bfseries\color{kipublue}}{\thesection.}{0.5em}{}
\titleformat{\subsection}{\large\bfseries\color{kipublue}}{\thesubsection}{0.5em}{}

\begin{document}

\begin{titlepage}
\vspace*{2cm}
\begin{flushleft}
{\color{kipuaccent}\rule{\linewidth}{2pt}}\\[1.5em]
{\Huge\bfseries\color{kipublue} 
Off-line quantum-advantage feature extraction for industrial production
}\\[1em]
{\Large\color{kipublue} How quantum feature surrogates turn expensive online quantum computation into an off-line deployable asset for real businesses}\\[1.5em]
{\color{kipuaccent}\rule{\linewidth}{1pt}}\\[1.5em]
{\normalsize Carlos Flores-Garrig\'os, Gabriel D. Alvarado Barrios, Qi Zhang, Anton Simen, Enrique Solano}\\[1.0em]
{\small Kipu Quantum GmbH, Greifswalderstrasse 212, 10405 Berlin, Germany}\\[3em]
\end{flushleft}
\vfill

\end{titlepage}

\tableofcontents

\newpage

\section{Abstract}

Quantum computing is no longer a lab curiosity for academic research. Industrial processors exceeding 100 qubits are commercially accessible and, for the first time, can extract information from data in ways that classical algorithms struggle to match. The most direct way to monetize this capability for industrial production today is \emph{quantum feature extraction}: turning raw business data (images, customer records, molecules, or sensor readings) into richer representations that outperform standard machine learning models.

There is one obstacle, however, that stands between today's demonstrations and tomorrow's production systems: \textbf{every sample of data costs a quantum computing execution}. For a company with millions of customers, satellite images, or transactions per month, processing every sample on quantum hardware is simply not viable.

This work introduces \textbf{quantum feature surrogates}, a framework developed by Kipu Quantum that breaks this bottleneck. The idea is intuitive though challenging: instead of asking the quantum computer to look at every single sample, we let it look at a small, carefully chosen subsample of the data, whose distribution faithfully represents the full set. A simple classical model, a \emph{surrogate}, then learns the quantum-induced patterns and applies them to the rest of the dataset at near-zero cost. The quantum processor stops being a per-sample engine and becomes a \emph{teacher of representations}, while production inference runs entirely on classical hardware.

\begin{takeaway}
\begin{itemize}[leftmargin=*,nosep]
\item \textbf{At least $\mathbf{5 \times}$ fewer quantum executions} for the same accuracy, and substantially more reduction as data volumes grow. Only a small, representative subsample of the data is processed quantumly.
\item \textbf{Classical inference} at deployment: no quantum queue, no per-prediction latency penalty.
\item \textbf{Same accuracy as the full quantum baseline}: confirmed on a real satellite-image benchmark, matching the full quantum pipeline.
\item \textbf{Industry-ready use cases}: satellite image classification, customer analytics, medical imaging, drug screening, churn prediction, and many more.
\end{itemize}
\end{takeaway}

\begin{figure}[t]
    \centering
    \includegraphics[width=\columnwidth]{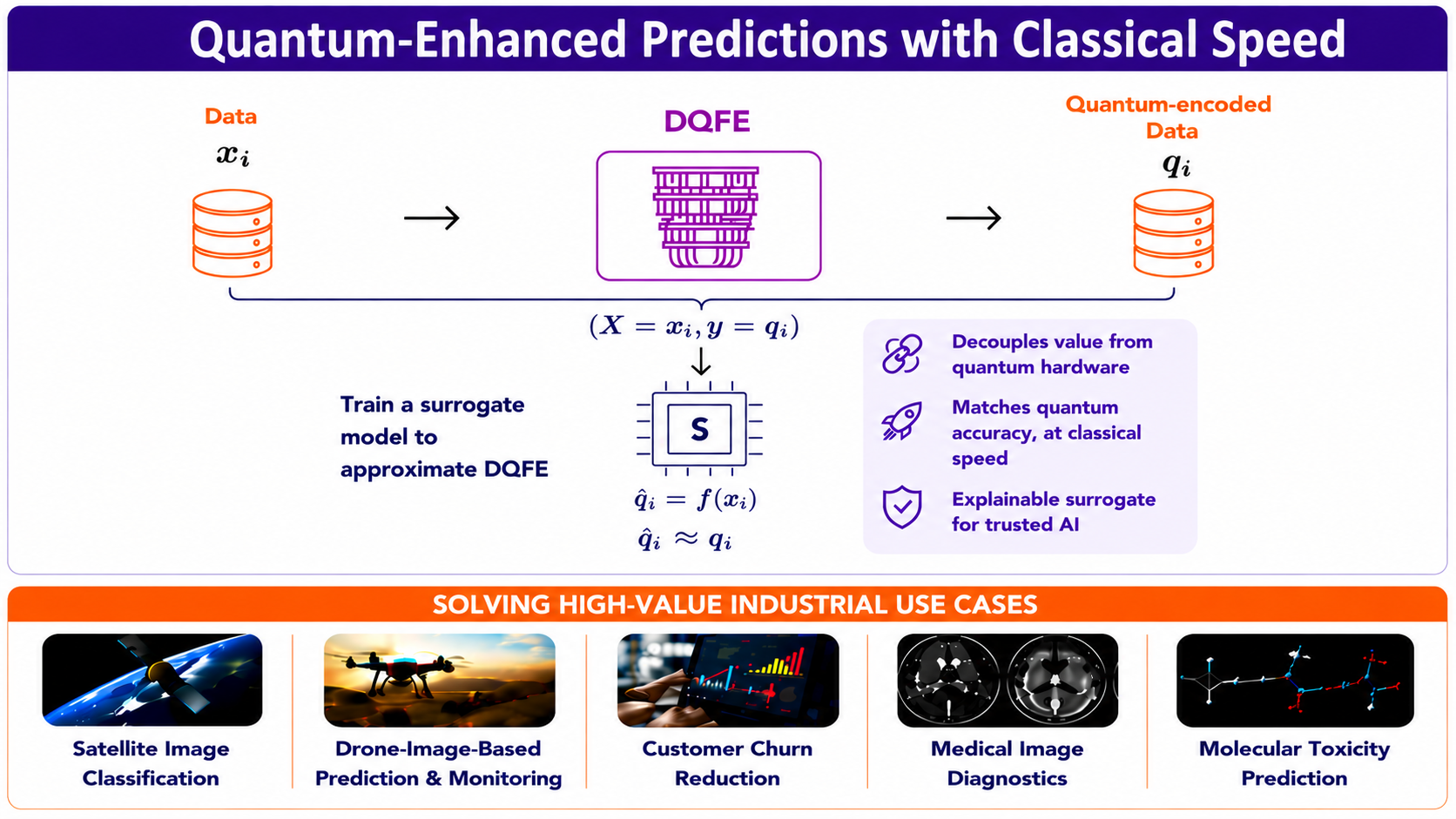}
    \caption{
    Off-line quantum-advantage feature extraction brings quantum-enhanced machine learning to real-world applications by combining quantum feature generation with fast, scalable classical inference, delivering measurable business value in seconds.
    }
    \label{fig:classical_distr_kmax7}
\end{figure}

\section{The Challenge at Industrial Scale}

Quantum computers can encode classical data into rich, non-linear representations that reveal patterns invisible to standard machine-learning tools. Kipu Quantum has achieved quantum-advantage evidence~\cite{simen2025digitizedcounterdiabaticquantumfeature} and demonstrated this on real-world problems, including image classification~\cite{simen2024digital}, molecular toxicity and medical imaging~\cite{simen2025quenchedquantumfeaturemaps}, and satellite image classification~\cite{zhang2026qsatellite}, with consistent $2$--$3\%$ absolute accuracy gains over strong classical baselines on IBM hardware, see Kipu's Blog post~\cite{kipu2025surrogates}.

The catch is operational. Every time you want a quantum-enhanced representation of a data point, the quantum computer must prepare a fresh state, evolve it through a tailored circuit, and measure it hundreds of times (from a queue shared with other users). For a research dataset of a few hundred samples, this is fine. For an enterprise dataset of millions, it is a non-starter. The cost grows \emph{linearly} with the number of samples, while business demand grows \emph{exponentially} as products go live.

The deeper observation is that none of this is a fundamental property of quantum-enhanced \emph{representations}. It is a property of how those representations are currently \emph{computed}. The geometry that quantum dynamics induce on the data is smooth, structured, and largely predictable from the original inputs. If we can capture that geometry once, we can replay it cheaply forever, and that is exactly what classical-surrogate research has begun to validate~\cite{schreiber2023classical,jerbi2024shadows}.

\section{The Solution: Quantum Feature Surrogates}

Kipu Quantum's framework rearranges the work so that the quantum computer is used where it is uniquely valuable, and classical hardware does everything else. The pipeline has five steps:

\begin{enumerate}[leftmargin=*]
\item \textbf{Pick a smart subsample.} From your training data, select a small subsample \emph{whose distribution is representative of the full dataset}: same class balance, same coverage of the input space, same edge cases. This is the single most important design choice. Stratified sampling combined with clustering ensures that minorities and rare patterns are not missed. The right subsample size depends on the problem and is typically a small fraction of the full dataset (larger volumes generally allow proportionally smaller subsamples).
\item \textbf{Run the quantum step once.} Process this subsample on quantum hardware using Kipu's Digitized Quantum Feature Extraction (DQFE)~\cite{simen2025quenchedquantumfeaturemaps,simen2024digital,Cadavid_2024,Cadavid_2025}. The output is a rich and non-linear representation of those samples.
\item \textbf{Train the surrogate.} A simple and regularized linear model (Ridge regression) learns to predict the quantum representation directly from the original input. Training takes seconds, no hyperparameter wars.
\item \textbf{Replay classically.} Apply the surrogate to the rest of the dataset (and to all future data). Every sample becomes a quantum-enhanced sample, computed as a single matrix multiplication.
\item \textbf{Train as usual.} Feed the resulting quantum-enhanced dataset to any classical ML model your team already uses: Random Forest, Gradient Boosting, Neural Networks, XGBoost, and others. No quantum expertise required downstream.
\end{enumerate}

The elegant part is that the surrogate does not have to ``do quantum operations''. The quantum computer has already produced the non-linear representation; the surrogate only needs to learn the \emph{geometry} of that representation, which by construction is smooth and well-behaved. When the geometry is more intricate, the same scheme extends to kernel-Ridge or shallow-network variants without changing the rest of the pipeline.

\section{Business Impact}

The framework's value translates directly into operational and financial metrics that executives recognize. The impact comes from four reinforcing axes: better models, faster inference, lower cost, and seamless integration with the infrastructure you already have.

\subsection*{Accuracy gain that survives deployment}

The quantum representation lifts classical baselines by a consistent $2$--$3\%$ points in absolute accuracy across Kipu Quantum's published benchmarks~\cite{simen2024digital,simen2025quenchedquantumfeaturemaps,zhang2026qsatellite}. On the KPMG satellite-imagery benchmark, this means $87\%$ accuracy versus $84\%$ from a strong ResNet-50 baseline (a gain that the surrogate preserves in full at a fraction of the quantum cost). In contexts such as fraud detection, churn prediction or medical image classification,  $2$--$3\%$ points of accuracy typically translate into millions of euros saved or thousands of additional positive outcomes per year, because the marginal value of every correctly classified sample is high.

\subsection*{Inference at millisecond latency}

Once the surrogate is trained, predicting on a new sample is a single matrix multiplication: of the order of \emph{milliseconds} on commodity CPUs, microseconds on GPUs. There is no quantum queue, no shot estimation, no calibration drift between calls. This unlocks the entire family of use cases that demand low-latency decisions (real-time fraud scoring, anomaly detection on sensor streams, customer intente) that a per-sample quantum pipeline simply cannot serve.

\subsection*{Off-line and batch predictions at industrial volumes}

The same property makes the framework ideal for \emph{off-line} and \emph{batch} workloads where millions of samples must be scored overnight: nightly credit-risk recalculations, periodic image re-classification of satellite archives, batch toxicity screening of compound libraries, monthly customer-segmentation refreshes. Because the surrogate runs on standard classical infrastructure, throughput scales linearly with the cluster you already pay for, and predictions can happen on-premise without exposing any data to an external quantum service.

\subsection*{Lower cost and operational simplicity}

Quantum executions are the most expensive ingredient in any quantum-enhanced pipeline. Replacing the full-dataset quantum pass with a representative subsample reduces the hardware bill, shortens the queue time, and decouples your release cadence from quantum-vendor availability. Downstream, the surrogate's output is just a tabular dataset, so your existing MLOps stack consumes it with zero changes. No new infrastructure, no new tooling, no new hires with niche quantum skills.

\begin{table}[h]
\centering
\begin{tabular}{p{4cm}p{9cm}}
\toprule
\textbf{Dimension} & \textbf{What changes} \\
\midrule
Accuracy            & Preserved: for example. $87\%$ on the KPMG satellite benchmark, matching the full quantum pipeline and beating the $84\%$ classical baseline. \\
Quantum cost        & Drops substantially: only a small representative subsample needs a quantum run, while accuracy is preserved. \\
Inference latency   & Milliseconds per sample: a single matrix multiplication, no quantum queue. \\
Throughput          & Scales with classical infrastructure---millions of predictions per day, offline or online. \\
Deployment mode     & Online, batch and edge---all supported, all classical. \\
Integration         & Plain tabular output; existing MLOps stack consumes it with no changes. \\
Operational risk    & No dependency on real-time quantum hardware availability for production traffic. \\
Vendor risk         & Surrogate is hardware-agnostic; benefits transparently from each new generation of devices. \\
\bottomrule
\end{tabular}
\caption{Operational impact of deploying the Quantum Feature Surrogate framework.}
\end{table}

\begin{takeaway}
You are not buying a quantum computer or real-time quantum computations from Kipu Quantum. You are buying a representation that makes your existing models measurably better, served at classical cost and classical speed: online, offline, or at the edge.
\end{takeaway}

\section{Use Cases}

\subsection{Satellite and drone image classification}
\begin{table}[h]
\centering
\begin{tabular}{lc}
\toprule
\textbf{Approach} & \textbf{Accuracy} \\
\midrule
Classical baseline (ResNet-50 + RF, $120$ feat.) & $84.0\%$ \\
Full quantum pipeline (DQFE, $120$ qubits)        & $87.0\%$ \\
\textbf{Ridge Surrogate (this framework, $120$ qubits)} & $\mathbf{87\%}$ \\
\bottomrule
\end{tabular}
\caption{Surrogate vs.\ full quantum vs.\ classical on the TreeSatAI benchmark, use case selected by KPMG. Classical and full-quantum figures from Ref.~\cite{zhang2026qsatellite}.}
\end{table}

A multi-class tree-genus classification task built on the TreeSatAI benchmark, combining Sentinel-1 SAR, Sentinel-2 multispectral and high-resolution aerial imagery~\cite{zhang2026qsatellite}. The dataset has $1{,}000$ training and $200$ test samples across five challenging classes. In this benchmark, the representative subsample passed to the quantum processor was $200$ samples which is one fifth of the training set ($20\%$, a $5\times$ reduction in quantum executions). The size of this subsample is benchmark-specific; in larger production settings (millions of samples), the same accuracy can be reached with a substantially smaller fraction.

\begin{figure}[t]
    \centering
    \includegraphics[width=\columnwidth]{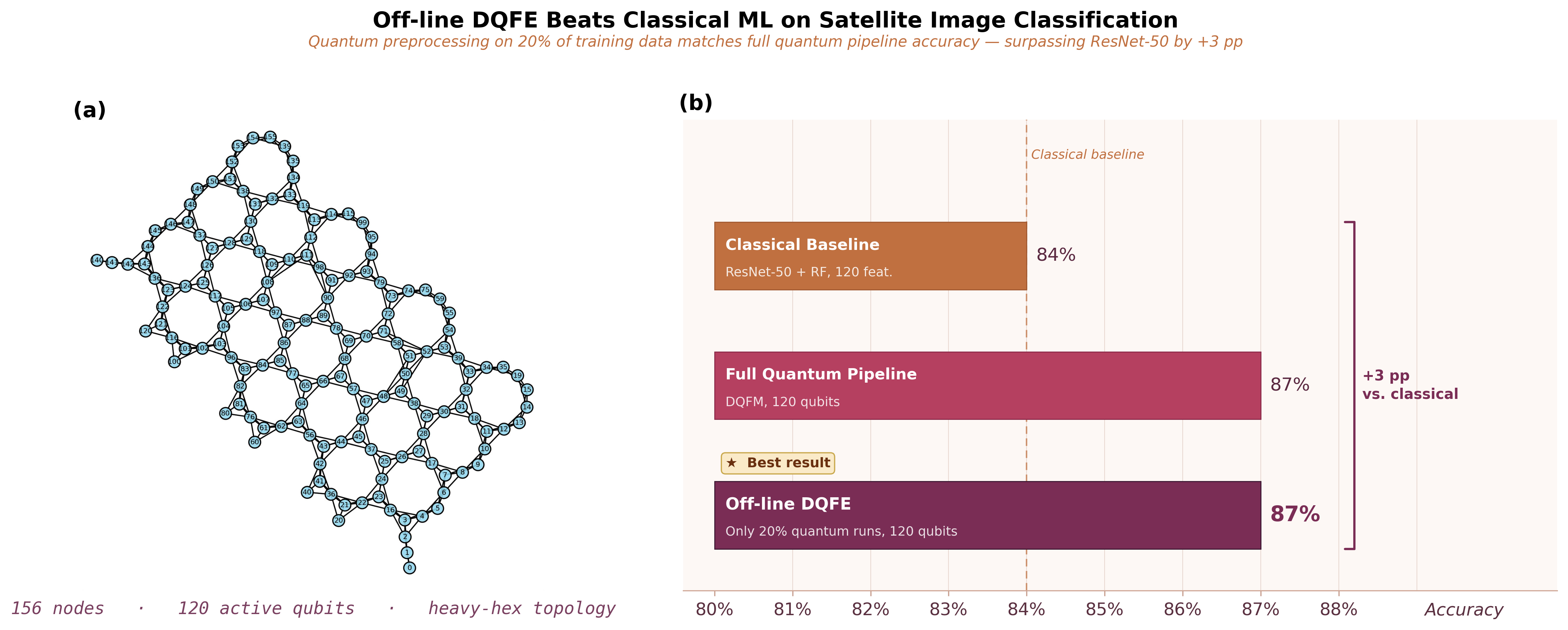}
    \caption{Quantum-enhanced tree-genus classification on the TreeSatAI benchmark.
\textbf{(a)} Coupling map of the IBM DQFE processor used for quantum feature encoding (156 nodes, 120 active qubits, heavy-hex topology).
\textbf{(b)} Classification accuracy of three approaches: a classical baseline (ResNet-50\,+\,RF), a full quantum pipeline (DQFE), and our Off-line DQFE, which matches the full quantum accuracy (87\%) at one fifth of the quantum cost, surpassing the classical baseline by 3\,pp. Classical and full-quantum results from~\cite{zhang2026qsatellite}.}
    \label{fig:quantum_classification}
\end{figure}

\textbf{Business meaning.}
Fig.~\ref{fig:quantum_classification} summarises our key result: by off-loading quantum feature extraction to a preprocessing stage on IBM's 120-qubit DQFE processor, Off-line DQFE matches the accuracy of a fully quantum pipeline (87\%) while reducing the number of required quantum runs by a factor of 5 or more, delivering the full benefit of the quantum-enhanced model at one fifth of the quantum cost. The customer, therefore obtains state-of-the-art accuracy on operational satellite pipelines, with predictions served at classical latencies and quantum hardware engaged only during the offline training phase.

\subsection{High-Volume Classification in Regulated Enterprise Settings}

A broad class of enterprise machine-learning problems shares the same structural profile: high-cardinality multi-class classification over heterogeneous features (text, numerical, behavioural), deployed at scale in regulated environments, and already served by mature classical baselines such as gradient-boosted trees or deep networks. Representative examples include customer-intent routing in contact centers, fraud and risk triage, claims and ticket classification, and transaction categorization. In all of these, years of feature engineering have compressed the remaining accuracy headroom, yet each additional percentage point of performance translates into material operational savings.

Two properties make this family of problems a natural fit for the classical surrogate framework. First, the data exhibits the kind of subtle, non-linear, long-tailed structure where richer quantum feature maps have shown consistent $2$--$3\%$ absolute accuracy gains in our published benchmarks across imaging, molecular, and remote-sensing domains. Second, production volumes routinely reach millions of samples per month, which rules out per-sample quantum processing on both cost and latency grounds. The surrogate approach resolves this tension directly: quantum hardware is invoked once on a distribution-preserving subsample, and the learned surrogate maps every production sample into the quantum-enhanced representation via a single matrix multiplication. The downstream classifier remains a standard classical model, fully compatible with existing MLOps, monitoring, governance, and explainability stacks.

Preliminary internal benchmarks on public enterprise-classification datasets (including customer churn and tabular classification tasks from open repositories) show accuracy uplifts consistent with the $2$--$3\%$ range observed in our peer-reviewed work.

This combination, accuracy lift where classical methods have plateaued, no quantum dependency at inference time, and seamless integration with existing production pipelines, makes the surrogate framework a particularly promising tool for the next generation of enterprise classification systems.

\subsection{Healthcare and life sciences}

The same recipe transfers to molecular toxicity prediction and medical-image diagnostics. On the Breast MedMNIST benchmark, Kipu's underlying quantum feature extractor outperforms deep-learning baselines such as ResNet-18 and ResNet-50~\cite{simen2025quenchedquantumfeaturemaps}. The Ridge surrogate brings this advantage into a deployable form for high-throughput screening pipelines while preserving virtually all of the quantum lift.

\begin{table}[h]
\centering
\begin{tabular}{lc}
\toprule
\textbf{Approach} & \textbf{AUC} \\
\midrule
ResNet-50                                       & $0.866$ \\
ResNet-18                                       & $0.891$ \\
Full quantum pipeline (DQFM)~\cite{simen2025quenchedquantumfeaturemaps} & $0.937$ \\
\textbf{Ridge Surrogate (this framework)}        & $\mathbf{0.932}$ \\
\bottomrule
\end{tabular}
\caption{Surrogate vs.\ full quantum vs.\ classical deep-learning baselines on the Breast MedMNIST benchmark. Classical and full-quantum figures from Ref.~\cite{simen2025quenchedquantumfeaturemaps}.}
\end{table}

\textbf{Business meaning.} The surrogate matches the full quantum pipeline within $0.005$ AUC while running on standard classical infrastructure---compatible with high-throughput screening and routine clinical-grade workflows.

\subsection{Other natural fits}
Among others, we can confidently mention insurance fraud detection and churn modeling, retail and banking recommendation and credit-risk scoring, predictive maintenance from manufacturing sensors, demand forecasting and anomaly detection on energy grids. Wherever data volumes are large and classical baselines are already strong, the surrogate framework is a candidate.

\section{Risks and Recommendations}

We are transparent about where the framework requires care. \textbf{The single most important design choice is the subsample}. Its distribution must faithfully represent the full datase: same class balance, same coverage of the input space, and same edge cases. When this is done well, the surrogate is robust; when the subsample is biased or sparse, it will extrapolate poorly. Inputs that fall outside the regions covered by the subsample should be flagged and routed back to the quantum processor, which is handled by active-learning and uncertainty-aware extensions. Highly non-smooth quantum representations may also underfit a linear surrogate; kernel-Ridge and shallow-network variants are drop-in replacements with the same training cost. Quantum hardware itself is still maturing, but the framework is hardware-agnostic and benefits transparently from each new generation of devices.

\begin{takeaway}
\textbf{For decision makers and executives, the recommendation is straightforward.} Identify one high-impact ML use case where (a) accuracy gains translate clearly into added business value, (b) data volumes preclude per-sample quantum runs, and (c) you already have a strong classical baseline. Run a pilot on that use case with a small representative subsample of your training data, measure the lift, and only then expand. The risk is bounded, the upside is measurable, and the deployment runs on the infrastructure you already own.
\end{takeaway}

\paragraph{Next steps.} Kipu Quantum offers an end-to-end engagement model that covers feasibility scoping, quantum hardware execution on IBM backends, surrogate training, and production handover to your existing data-science teams. To start a conversation, contact us at \href{mailto:info@kipu-quantum.com}{info@kipu-quantum.com} or visit \href{https://kipu-quantum.com}{kipu-quantum.com}.

\appendix
\section{Appendix}\label{app:methodology}

This appendix complements the business-level description of the framework with the technical formulation that underlies it. The full derivations and additional details can be found in this technical section.

Let $\mathcal{D}_{\text{train}}=\{(\mathbf{x}_i,y_i)\}_{i=1}^{N}\!\subset\!\mathbb{R}^{d}\!\times\!\mathcal{Y}$ be the training set and $\Phi:\mathbb{R}^{d}\!\rightarrow\!\mathbb{R}^{D}$ the (in general non-linear) quantum feature map whose direct evaluation on every sample is not affordable. The framework rearranges this evaluation so that the quantum-induced geometry is preserved while the cost becomes largely independent of $N$. A subsample $\mathcal{D}_Q\!\subset\!\mathcal{D}_{\text{train}}$ of size $M\!\ll\!N$ is processed quantumly to produce the pairs $\{(\mathbf{x}_i,\Phi(\mathbf{x}_i))\}_{i=1}^{M}$, a regularized linear map $F_{\boldsymbol{\theta}}$ is fitted on those pairs, and the remaining samples are transformed classically through $\hat{\mathbf{x}}_j=F_{\boldsymbol{\theta}}(\mathbf{x}_j)$. At deployment, a new sample $\mathbf{x}^{\star}$ is pushed through $F_{\boldsymbol{\theta}}$ and the downstream model with no further quantum executions. The three pieces that fully determine the behaviour of this pipeline are described next.

\subsection{Subsample selection}

The surrogate can only be trusted in regions of the input space where the quantum signal has actually been measured, so the choice of $\mathcal{D}_Q$ defines its operational domain. We combine stratified sampling on the labels with $k$-medoid clustering in the input space: stratification protects minority classes, while clustering ensures that medoids cover the regions the data actually populate. Alternatives such as low-discrepancy sequences, diversity-based scores or active-learning rules that grow $\mathcal{D}_Q$ on demand can replace this step without affecting the rest of the pipeline.

\subsection{Quantum feature extraction}

The framework is agnostic to the specific embedding, but it is most naturally paired with the Hamiltonian-based feature extractors recently demonstrated on $156$-qubit IBM hardware~\cite{simen2025quenchedquantumfeaturemaps,simen2024digital,zhang2026qsatellite}. A sample $\mathbf{x}$ is encoded into the longitudinal fields and higher-order couplings of a $k$-local spin-glass Hamiltonian,
\begin{equation}\label{eq:Happ}
  H(\mathbf{x})=\sum_{i=1}^{n}x_i\,\sigma_i^{z}+\sum_{k=2}^{K}\sum_{S\in\mathcal{G}^{(k)}}\!c_S\!\prod_{i\in S}\sigma_i^{z},
\end{equation}
where $\mathcal{G}^{(k)}$ is the hypergraph of $k$-body interactions allowed by the device connectivity and $c_S$ are higher-order mutual-information coefficients estimated from the dataset. The system is driven by an adiabatic Hamiltonian $H_{\text{ad}}(t;\mathbf{x})=A(t)H_i+B(t)H(\mathbf{x})$ augmented with a first-order counterdiabatic term and Trotterized in the impulse regime~\cite{Cadavid_2024,Cadavid_2025}, which suppresses non-adiabatic transitions while keeping the circuit shallow. The output features $\Phi(\mathbf{x})$ are collected as a vector of single- and higher-body expectation values,
\begin{equation}\label{eq:phiapp}
  \Phi(\mathbf{x})=\sum_{i=1}^{n}\!\langle\sigma_i^{z}\rangle\,\mathbf{e}_i+\sum_{k=2}^{K}\!\sum_{S\in\mathcal{G}^{(k)}}\!\Big\langle\!\prod_{i\in S}\!\sigma_i^{z}\Big\rangle\mathbf{e}_S,
\end{equation}
acting as a correlation-aware, non-linear representation of $\mathbf{x}$. The cost of producing this representation is paid only on $\mathcal{D}_Q$.

\subsection{Ridge surrogate}

The surrogate is a regularized affine map $F_{\boldsymbol{\theta}}(\mathbf{x})=W\mathbf{x}+\mathbf{b}$ with parameters $\boldsymbol{\theta}=(W,\mathbf{b})$, fitted by minimizing
\begin{equation}\label{eq:lossapp}
  \mathcal{L}(\boldsymbol{\theta})=\tfrac{1}{M}\!\sum_{i=1}^{M}\!\big\|\Phi(\mathbf{x}_i)-F_{\boldsymbol{\theta}}(\mathbf{x}_i)\big\|_2^{2}+\lambda\,\|W\|_F^{2},
\end{equation}
where $\lambda>0$ controls the bias--variance trade-off and admits a closed-form solution that makes training stable, hyperparameter-light and reproducible. When the data manifold shows residual non-linearity that a purely linear map cannot absorb, the same scheme accepts random Fourier features or low-degree polynomial lifts of $\mathbf{x}$, smoothly bridging towards kernel-Ridge variants~\cite{schreiber2023classical,jerbi2024shadows}.

The conceptual point is that the surrogate does not learn quantum mechanics: Eq.~\eqref{eq:phiapp} is already deeply non-linear in $\mathbf{x}$ through the unitary dynamics, and the surrogate only needs to interpolate the smooth observable manifold induced by Eq.~\eqref{eq:Happ}. The quantum processor produces the geometry; the Ridge map carries it, almost for free, to the rest of the dataset.

\bibliography{bibfile}

\end{document}